\documentclass[reprint,amsmath,amssymb,aps,]{revtex4-2}
\usepackage{graphicx}
\usepackage{dcolumn}
\usepackage{bm}
\usepackage{lipsum}
\usepackage[colorlinks=true,breaklinks=true,linkcolor=blue,
            citecolor=blue,urlcolor=blue]{hyperref}
\usepackage{tgtermes}
\usepackage{txfonts}
\usepackage{xcolor}
\usepackage{asymptote}
\usepackage{tikz}

\newcommand{\defeq}{\overset{\underset{\mathrm{def}}{}}{=}}

\renewcommand{\vec}[1]{%
  \ifcat\noexpand#1\relax 
    \boldsymbol{#1} 
  \else
    \mathbf{#1} 
  \fi
}

\newcommand{\uvec}[1]{
  \ifcat\noexpand#1\relax 
    \hat{\boldsymbol{#1}} 
  \else
    \hat{\mathbf{#1}} 
  \fi
}

\begin{document}

\preprint{APS/123-QED}

\title{Light drag in nonuniformly moving anisotropic media \\ through the lens of gradient-index optics}

\author{Julien Langlois}
\email{julien.langlois@laplace.univ-tlse.fr}
\author{Renaud Gueroult}
\affiliation{LAPLACE, Université de Toulouse, CNRS, INPT, UPS, 31062 Toulouse, France}

\date{\today}

\begin{abstract}
    The trajectory of light rays propagating through a nonuniformly moving anisotropic medium is determined by considering the Fresnel drag experienced by the wave at each point along the ray. By showing that symmetries in the velocity field manifest as symmetries in the effective wave index representing the moving medium, methods classically employed to model gradient index media are then used to obtain analytical forms for the ray trajectory. When applied to isotropic media, the results are verified to be consistent with those obtained using an optical (Gordon) metric. The potential of this method to model light rays in anisotropic media is finally demonstrated by considering waves in a nonuniformly moving magnetized plasma, exposing how nonuniform motion and anisotropy can compete with one another.   
\end{abstract}

\maketitle

\section{Introduction} \label{sec:intro}

As first postulated by Fresnel~\cite{fresnel1818} and subsequently demonstrated by Fizeau's 1851 experiment~\cite{fizeau1860xxxii}, the propagation of light in a medium is affected by the medium’s motion. A wave propagating along a uniformly-moving medium appears to have a sped-up or slowed-down phase velocity, depending on whether the wave and the medium have identical or opposite directions. Lorentz pointed out that dispersion should in fact contribute to this longitudinal light drag~\cite{lorentz1895versuch}, which was later confirmed experimentally by Zeeman~\cite{zeeman1914fresnel,zeeman1915fresnel}, firmly establishing the drag coefficient 
\begin{equation}
    f=1-\frac{1}{n n_g}
\end{equation}
with $n$ and $n_g$ the phase and group index of the medium, respectively. Considering this time a motion transverse to the wavevector, Jones demonstrated that the medium's motion leads in this case to a transverse deflection of the wave beam~\cite{jones1972fresnel,jones1975}. Because this lateral drag depends on the same drag coefficient $f$~\cite{player1975,rogers1975}, it is now referred to as Fresnel's transverse drag. These light-dragging phenomena can be explained by considering propagation in the frame in which the medium is at rest, with the help of Lorentz transformations~\cite{player1975,ko1978passage,pauli2013theory,carusotto2003transverse}.

Although the discovery of light-dragging effects proved to be an important step in the development of relativity~\cite{whittaker1989history,Cassini2019}, the fact that these effects are small in typical dielectrics for nonrelativistic velocities~\cite{landau2013electrodynamics} initially limited practical implications. On the other hand, the development and generalization of slow-light media featuring artificially large group index $n_g$~\cite{kash1999ultraslow,franke2011rotary,artoni2001fresnel,safari2016light,kuan2016large,qin2020fast}, and the discovery of exotic media such Bose-Einstein condensates~\cite{matthews1999vortices}, changed this standing. In fact, significant light-dragging effects are now routinely observed in stimulated media~\cite{hau1999light,fleischhauer2005electromagnetically,bigelow2003observation}, and have been predicted~\cite{Gueroult2019,Gueroult2020,Rax2021,Rax2023b,Gueroult2023} and recently observed~\cite{Gueroult2025} in plasmas. 

Practical applications, however, often involve nonuniform velocity fields. This was in fact already true of Jones' demonstration of transverse drag~\cite{jones1972fresnel,jones1975} as he used a rotating disk, but numerous applications of waves involve propagation in a nonuniformly-moving media. This is for instance the case of the RF waves used to heat and drive current in a tokamak plasma for magnetic confinement fusion~\cite{Fisch1987}, or of the satellite signal propagating in a rotating atmosphere in atmospheric occultations studies~\cite{bourgoin2020general}.

Surprisingly, while the effect of a nonuniform motion on acoustic rays had already gained interest during the second half of the twentieth century (see, e.g.,~\cite{kornhauser1953ray,uginvcius1972ray,thompson1972ray}), this same problem but for light rays appears to have attracted much less attention until Leonhardt and Piwnicki's work in the early 2000s~\cite{leonhardt1999optics,leonhardt2000relativistic,leonhardt2001slow}. Indeed, while earlier contributions did consider the problem of an accelerated motion~ \cite{lerche1974passage,censor1976ray,tanaka1982reflection}, the focus was then largely on addressing the fundamental question of the consequences of acceleration on electromagnetic field equations and on the associated constitutive relations~\cite{Heer1964,Anderson1969,shiozawa1973phenomenological}, rather than on the effect on the wave trajectory.

In their original paper~\cite{leonhardt1999optics}, Leonhardt and Piwnicki develop a relativistic theory of light propagation in nondispersive isotropic media. It uses the classical ray tracing equations but for motion-corrected optical Hamiltonian and Lagrangian. These operators are derived from an optical metric, exploiting the profound analogy between gravity and moving dielectrics discovered by Gordon~\cite{gordon1923lichtfortpflanzung}. In essence, the use of this metric amounts to applying instantaneous Lorentz boosts along the optical path, considering that the characteristic length scale of the velocity field is large compared to the wavelength~\cite{censor1976ray}. In fact, Balazs had already considered in 1955 the possibility of using Gordon's analogy to sketch a generalization of Fermat's principle to moving bodies~\cite{balazs1955propagation}. Considering then dispersion effects, Leonhardt and Piwnicki proposed to modify Gordon's metric to capture the response of dispersive isotropic media near  cutoffs~\cite{leonhardt2000relativistic,leonhardt2001slow}, but a general effective curvature valid for any dispersive isotropic media is still missing. \textit{A fortiori}, the extension of this covariant approach to anisotropic media remains an open problem.

For more general moving media, including dispersive anisotropic media such as plasmas, the optical path could in principle be obtained by taking advantage the covariance of dispersion relations~\cite{censor1980dispersion,melrose1973covariant} to construct an effective Hamiltonian~\cite{braud2025ray,Braud2025a}. This approach generalizes the theory proposed by Walker considering only the effect of the Doppler shift~\cite{Walker2008}. While convenient and promising, working with the dispersion relations of this effective medium conceals in some ways the physics at play. In this paper, we propose instead to exploit another characteristic brought in by motion. Indeed, while motion induces as mentioned above an effective anisotropy, the nonuniformity of the velocity field also introduces an effective inhomogeneity. A moving medium can thus be examined through the prism of gradient-index (GRIN) optics~\cite{born2013principles,merchand2012gradient}. Since the effective Hamiltonian is built upon an instantaneous Lorentz transformation, deriving the optical path boils down, as we will show, to computing a series of instantaneous uniform transverse drags~\cite{player1975,ko1978passage}. The approach proposed here can hence be regarded as a relativistic analogue to the historical method used in GRIN dielectrics, that is discretizing the medium into infinitesimal homogeneous layers and applying the laws of refraction between each layer~\cite{broer2014bernoulli,trager2012springer}.

This paper is organized as follows. First, we show in Sec.~\ref{sec:local_drags} how the recently derived theory of drag for anisotropic media in uniform linear motion~\cite{langlois2025fresnel} can be extended to a generic velocity field. Then, we demonstrate in Sec.~\ref{sec:sym} that the existence of symmetries in the velocity field, just like symmetries in the wave index in standard GRIN media, makes it possible to simplify significantly the ray trajectory equation. Taking advantage of this analogy, we derive in Sec.~\ref{sec:app} analytical solutions for the ray trajectory for both boost and rotation, recovering results known for isotropic dielectrics, but also exposing new results that may prove useful to validate simulation tools. We finally demonstrate in Section~\ref{sec:plasma} the potential of the method for anisotropic medium, exposing the combined effect of anisotropy and drag in a rotating magnetized plasma. Lastly, Sec.~\ref{sec:conclu} summarizes the main findings of this study. 

\section{Light path as the limit of infinitesimal local uniform drags} \label{sec:local_drags}

\subsection{Refraction laws for a dispersive  anisotropic medium in uniform linear motion} \label{subsec:uniform}

Light-dragging phenomena are nowadays generally understood in terms of the theory of special relativity. Specifically, the classical laws of refraction are applied in the comoving-frame $\Sigma'$ where the medium is instantaneously stationary, whereas Lorentz velocity transformations~\cite{einstein1905elektrodynamik,landau2013classical} are used to rewrite these relations in terms of laboratory observables. This method has historically been proposed and used for isotropic dielectrics~\cite{player1975,ko1978passage}, for which one can take advantage of the fact that the phase velocity and the group velocity as seen in the comoving-frame are aligned. Yet, it has recently been shown that this approach can be extended to anisotropic media, under the condition that the dispersion relation is known in the comoving-frame~\cite{langlois2025fresnel}. We briefly here sum up the key findings of this work as it will serve as the foundation for the extension of drag to general velocity fields.

Consider an anisotropic body in uniform linear motion with constant velocity $\vec{v}$ in the laboratory-frame $\Sigma$. We write ${\vec{\beta}=\vec{v}/c}$ the normalized velocity and ${\gamma=(1-\beta^2)^{-1/2}}$ the Lorentz factor. Quantities expressed in the comoving-frame (or rest-frame) $\Sigma'$ are identified with primes. Quantities relative to the medium at rest---i.e., absent motion---are identified with a bar. Let $\mathcal D'$ be the dispersion relation of a given eigenmode supported by the medium in the comoving-frame, defined by
\begin{equation} \label{eq:dispersion_relation}
    \mathcal D' \defeq n'\left(\omega',\uvec{k}'\right)-ck'/\omega' = 0
\end{equation}
where the rest-frame optical index $n'$ is in general function of both the frequency $\omega'$ and the direction of wavevector ${\uvec{k}'=\vec{k}'/k'}$. Note that here ${n'=\bar{n}}$ by virtue of Newton's first law. Indeed, because we consider a uniform linear motion, the rest-frame is also inertial, and as a result the eigenmodes of the medium in $\Sigma'$ are those observed in $\Sigma$ in the absence of motion. 
Writing $k_\parallel'$ and $k_\perp'$ the components of the wavevector that are respectively parallel and perpendicular to the velocity $\vec{v}$, the dispersion relation can be rewritten as an equation $k_\perp'$ in Eq.~\eqref{eq:dispersion_relation} as a function of $\omega'$ and $k_\parallel'$. We accordingly define
\begin{equation} \label{eq:N_prime}
    \mathcal K' \defeq k_\perp'(\omega',k_\parallel')
\end{equation}
such that
\begin{equation}    
    \quad \mathcal D'(\omega',k_\parallel',\mathcal K')=0.
\end{equation}
The rationale for this rewriting, as it will become clear in the next paragraph, is that in the case of an interface that is parallel to the motion, the parallel wavevector $k_\parallel'$ is conserved, so that $k_\parallel'$ is simply set by the property of the incident wave.

\begin{figure}
    \centering
    \includegraphics{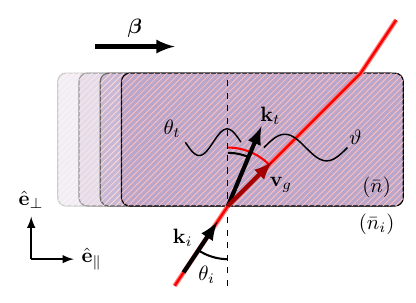}
    \caption{Sketch of the light-dragging effect observed in lab-frame as a result of propagation in a uniformly-moving (anisotropic) dielectric. The incidence angle, defined with respect to the velocity normal, is $\theta_i$. The refracted angle is $\theta_t$, whereas the refracted ray direction is $\vartheta$.}
    \label{fig:drag_scheme}
\end{figure}

Now consider that this medium is of finite extent, and that crucially the interface is parallel to the direction of motion. For an incident wave with a lab-frame wavevector $\mathbf{k}_i=k_i(\sin\theta_i\mathbf{\hat{e}}_{\parallel}+ \cos\theta_i\mathbf{\hat{e}}_{\perp})$ as shown in Fig.~\ref{fig:drag_scheme}, one can show that the angles of the transmitted wavevector $\theta_t$ and ray velocity $\vartheta$ relative to the direction normal to the motion write
\begin{subequations} \label{eq:uniform_refraction_laws}
\begin{equation} \label{eq:generalized_snell}
    \tan \theta_{t} = \frac{\omega}{c}\frac{\bar n_i\sin \theta_i}{\mathcal K'(\omega', k_{\parallel}')}
\end{equation}
and
\begin{equation} \label{eq:Psi}
    \tan \vartheta = \gamma\Bigg[c\beta \frac{\partial \mathcal K'}{\partial \omega'}(\omega', k_{\parallel}') - \frac{\partial \mathcal K'}{\partial k_{\parallel}'}(\omega', k_{\parallel}')\Bigg]
\end{equation}
with
\begin{align} \label{eq:evalutated_at1}
    \omega' &= (1-\beta \bar n_i \sin \theta_i)\gamma\omega, \\ \label{eq:evalutated_at2}
    k_{\parallel}'&=(\bar n_i\sin \theta_i-\beta)\gamma\omega/c.
\end{align}
\end{subequations}
This derivation is precisely that given in Ref.~\cite{langlois2025fresnel}, other than for the fact that we consider here the possibility for the incident wave to be in a medium, and simply write $k_i=\bar{n}_i\omega/c$ with $\bar{n}_i$ the wave index for the particular mode considered. Note importantly that Eq.~\eqref{eq:generalized_snell} and Eq.~\eqref{eq:Psi} involve the primed frequency and optical index as given in Eqs.~\eqref{eq:evalutated_at1} and \eqref{eq:evalutated_at2}. This captures the relativistic Doppler shift and aberration effect between the two frames~\cite{einstein1905elektrodynamik}.

While Eq.~\eqref{eq:Psi}, by relating the transmitted group velocity $\vec{v}_g$ to the incident wavevector $\vec{k}_i$, constitutes the law of refraction of the beam (in that it characterizes the direction of energy in wave packets~\cite{born2013principles}), Eq.~\eqref{eq:generalized_snell} actually provides the generalization of Snell's law. In fact, this equation can be recast in the more standard form~\cite{langlois2025fresnel}
\begin{equation} \label{eq:generalized_snell2}
    n\sin\theta_t = \bar n_i\sin\theta_i
\end{equation}
where $n$ is the laboratory-frame's phase index for the considered eigenmode, that is the refractive index which captures the anisotropy that results from the preferred direction introduced
by the motion. An observer in $\Sigma$ may in fact consider the moving medium as a 'stationary' medium, with an effective refractive index $n$ that depends both on the optical properties of the medium in $\Sigma'$ and on the velocity $\vec{v}$. This way the optical path can be obtained as in any static medium~\cite{cheng1968covariant,lopez1996dispersion,gjurchinovski2007fermat}. It is worth noting, however, that such effective consideration must be regarded with caution as the one-to-one analogy with material anisotropy is not guaranteed. In particular, optical indices of a moving isotropic medium are shown to be degenerate~\cite{mccall2007relativity}---i.e. independent of the polarization directions of the fields---so that motion does not result in birefringence as observed in static axial crystals or magnetized plasmas. This property actually depends on the frame in which stationary refraction laws apply~\cite{deck2021electromagnetic}.

Note finally that we chose throughout this section for simplicity to have the incident wavevector $\vec{k}_i$ to be in the plane defined by the interface normal and $\vec{\beta}$, but that this analysis could straightforwardly be extended to a wavevector with a component out of this plane.

\subsection{Nonuniform motion through GRIN optics} \label{subsec:localGRIN}

A direct consequence of considering a moving medium as an effective medium at rest with motion-dependent properties, and notably as seen above a motion-dependent refractive index $n(\vec{\beta})$, is that a nonuniform velocity field $\vec{\beta}(\vec{x})$ leads to an inhomogeneous refractive index $n[\vec{\beta}(\vec{x})]$. Under this analogy it then becomes possible to approach ray tracing in a moving medium through the lens of gradient-index (GRIN) optics, that is the study of light propagation through inhomogeneous media.   

A historical contribution of GRIN optics is Maxwell's famous 1854 proposal of the fish-eye lens~\cite{maxwell1854cambridge,merchand2012gradient}. Yet, almost two centuries before, Bernoulli had already considered the effect of a gradient-index medium when he published in the \textit{Acta Eruditorum} his solution of the brachistochrone problem, determining the curve along which an object placed in a uniform gravity field must slide so that its travel time is minimal~\cite{bernoulli4johannis}. Indeed, Bernoulli actually solved the problem in terms of a light ray. Specifically, he discretized a medium with a continuous refractive index profile into a finite number of homogeneous layers in which light must travel in a straight line. Applying refracting laws between each layer and taking the limit of zero thickness, he then derived the solution invoking Fermat’s principle that the light should follow the path of least time~\cite{broer2014bernoulli}. 

\begin{figure}[b]
    \centering
    \includegraphics[width=1\linewidth]{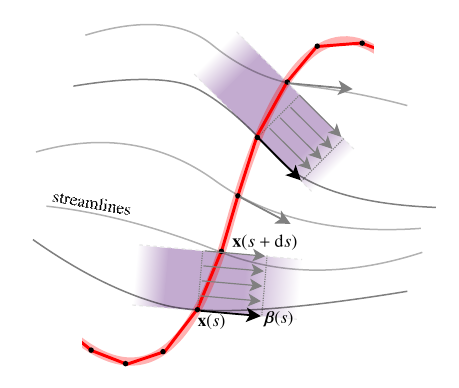}
    \caption{Model for the ray trajectory in a nonuniform velocity field. The trajectory is obtained as the limit of successive uniform drags at the interface between two infinitesimally thin layers with different but constant velocity. Since layers are homogeneous, light travels in straight line across each layer. This is analogous to the stratification employed for inhomogeneous refractive indexes in GRIN media.}
    \label{Fig:Layers}
\end{figure}

Here, we propose to similarly discretize the velocity profile representing a nonuniform motion into a series of homogeneous layers with constant velocity, and to apply at each interface between these layers the refraction laws at a moving interface recalled in Sec.~\ref{subsec:uniform}. This is illustrated in Fig.~\ref{Fig:Layers}. Consistent with ray theory, we demand that the characteristic properties of the effective medium, hence including the velocity, do not vary significantly over one wavelength and one period, respectively. The assumption of slow variations of the velocity also allows us to neglect acceleration effects~\cite{Heer1964,Anderson1969,shiozawa1973phenomenological,langlois2023contribution}, and thus assume special relativity theory to be valid locally and instantaneously~\cite{censor1976ray}. Under these hypotheses, the analog of GRIN optics for moving media entails modeling ray trajectory as the integral of instantaneous uniform drags along the optical path. Because this in essence corresponds to considering successive Lorentz boosts from one instantaneous rest-frame to another, all along the optical path, it is very much analogous to the covariant formalism associated with Gordon's optical metric~\cite{gordon1923lichtfortpflanzung}, and to the construction of optical operators as done by Leonhardt and Piwnicki~\cite{leonhardt1999optics}. In fact, we will show that we recover Leonhardt and Piwnicki's conclusions in Sec.~\ref{sec:app} when considering a nondispersive isotropic medium. We stress out that an advantage of the proposed method is that it applies equally to anisotropic media, as we will illustrate in Sec.~\ref{sec:plasma}. 

\subsection{Local refraction laws} \label{subsec:local}

As introduced above, we consider instantaneous uniform drags at the interface between two layers with different velocities. For this we apply the results of Sec.~\ref{subsec:uniform}, considering as inputs the wavevector and beam direction in the layer from which the ray emerges, as seen in the lab-frame $\Sigma$, and refraction of this ray on the moving layer on which the ray is incident. Introducing a ray parameter $s$, the relation for the change in ray direction Eq.~\eqref{eq:Psi} gives
\begin{subequations} \label{eq:nonuniform_Psi_full}
\begin{equation} \label{eq:nonuniform_Psi}
    \tan \vartheta(s) = \gamma\big[\mathbf x(s)\big]\Bigg\{c\beta\big[\mathbf x(s)\big] \frac{\partial \mathcal K'}{\partial \omega'}(\omega', k_{\parallel}') - \frac{\partial \mathcal K'}{\partial k_{\parallel}'}(\omega', k_{\parallel}')\Bigg\}
\end{equation}
where here
\begin{align} \label{eq:evaluated1}
    \omega' &= \Big\{1-\beta\big[\vec{x}(s)\big] n\big[\vec{x}(s)\big] \sin \theta(s)\Big\}\gamma\big[\vec{x}(s)\big] \omega, \\ \label{eq:evaluated2}
    k_{\parallel}'&=\Big\{n\big[\vec{x}(s)\big]\sin \theta(s)-\beta\big[\vec{x}(s)\big]\Big\}\gamma\big[\vec{x}(s)\big] \omega/c.       
\end{align}
\end{subequations} 
Note importantly that the angles $\vartheta$ and $\theta$ obtained here are, as shown in Fig.~\ref{fig:drag_scheme}, both defined with respect to the normal to the direction of motion $\boldsymbol \beta$, which is local and can vary along the trajectory. Note also that Eq.~\eqref{eq:nonuniform_Psi} is not in itself enough to compute the trajectory, as it depends as shown in Eqs.~\eqref{eq:evaluated1} and \eqref{eq:evaluated2} on the local incidence angle $\theta(s)$. This additional complexity compared to classical GRIN media stems from the fact that, since we consider here anisotropic media, the wavevector and the group velocity (i.e., the ray direction) are not necessarily aligned. An equation evolution for $\theta(s)$ could, however, be obtained in a similar fashion as in ray tracing in anisotropic media~\cite{braud2025ray}, which together with Eq.~\eqref{eq:nonuniform_Psi} would form a complete set of equations to be solved numerically to model the trajectory in a nonuniformly moving anisotropic media. Rather than going this route though, we choose here to exploit, in the spirit of GRIN optics studies, symmetry properties to expose analytical solutions. Specifically, symmetries in the refractive index are classically used in GRIN media to identify conserved quantities~\cite{born2013principles,gomez2021generalization}. Pursuing the analogy, we use here symmetries in the velocity field, which translates into symmetries in the effective phase index, to eliminate the need for the auxiliary equation for $\theta(s)$.

\section{Optical symmetries induced by motion} \label{sec:sym}

Having identified a system of coupled equations for the evolution of the ray trajectory and of the wavevector in a moving anisotropic medium, we now show how this system can be simplified to a single evolution equation for the ray trajectory using symmetry properties of the velocity field. 

\subsection{Fermat's invariant}

According to Noether's theorem, a symmetry of the Lagrangian defined from an action principle implies a conservation law on the associated Euler-Lagrange equations~\cite{kosmann2011noether}. In optics, this action principle corresponds to Fermat's principle~\cite{born2013principles}, stating that the optical length of the path followed by light between two points $a$ and $b$ is stationary. It is classically written in the form
\begin{equation} \label{eq:fermat_principle}
    \delta\int_a^b n\big[\vec{x}(s),\dot{\vec{x}}(s)\big]~{\rm d}s = 0
\end{equation}
where $n$ is the optical index of a mode supported by an anisotropic medium and ${\dot{\vec{x}}(s)={\rm d}\vec{x}/{\rm d}s}$ is the beam direction~\cite{rivera1995hamiltonian,holm2011geometric}. Note that by using $n$ in the integral, the physical meaning of the ray parameter~$s$ differs from that classically used for isotropic media. For $s$ to represent the same physical quantity, $n$ should be replaced by the ray index $n_r$. The latter is associated with the projection of the phase velocity onto the ray direction, which in an anisotropic medium is given by the group velocity~\cite{born2013principles,haselgrove1954ray}, as opposed to the phase velocity itself. Yet, this rewriting poses no issue as the integral is invariant under reparametrization of the optical path~\cite{holm2011geometric,vcerveny2002fermat}. By using this notation, the results derived here are applicable both to isotropic and anisotropic media, but one should keep in mind that an appropriate rescaling of the ray parameter is required.

One verifies that Eq.~\eqref{eq:fermat_principle} is in fact Hamilton's principle written for the optical Lagrangian
\begin{equation}
    \mathcal L\big[\vec{x}(s),\dot{\vec{x}}(s)\big] = n\big[\vec{x}(s),\dot{\vec{x}}(s)\big]|\dot{\vec{x}}(s)|
\end{equation}
that carries all the dynamical information of the system, from which Euler-Lagrange equations give
\begin{equation} \label{eq:euler_lagrange}
    \frac{\partial \mathcal L}{\partial \vec{x}}=\frac{\rm d}{\rm d s}\left( \frac{\partial \mathcal L}{\partial \dot{\vec{x}}}\right ).
\end{equation}
If the optical index---and thus the associated optical Lagrangian---does not depend on a certain coordinate $x_i$, Eq.~\eqref{eq:euler_lagrange} then directly gives that
\begin{equation}
    \frac{\partial \mathcal L}{\partial \dot x_i}=\mathrm{cst}\defeq\mathfrak F.
\end{equation}
The constant ${\mathfrak F\in\mathbb R}$ is the conserved quantity predicted by Noether's theorem. This constant, sometimes referred to as Fermat's invariant, is essentially an infinitesimal version of Snell's refraction law for symmetric GRIN media~\cite{gomez2021generalization}. 

\subsection{Conformal mapping}

Now, consider a transformation ${\vec{x}\to \vec{X}}$ which leaves the Lagrangian variationally invariant, that is to say that the variational principle Eq.~\eqref{eq:fermat_principle} can be rewritten as
\begin{equation} 
    \delta\int_A^B N\big[\vec{X}(S),\dot{\vec{X}}(S)\big]~{\rm d} S = 0,
\end{equation}
with $A,B,N,\dot{\vec X},S$ the transformed $a,b,n,\dot{\vec x},s$ respectively. If the transformed optical index $N$ does not depend on a certain $\vec X$ coordinate, then it is always possible to determine a Fermat's invariant for the original coordinates system $\vec x$. Naturally, it remains to be determined how to find such a transformation. A possibility in two dimensions, as explored in a number of studies~\cite{leonhardt2006optical,luneburg1966mathematical,coleman2004generalization}, is to use conformal mapping methods. 

\begin{figure*}
    \centering
    \includegraphics{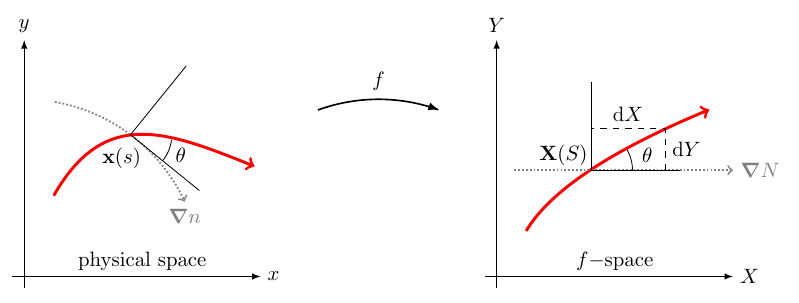}
    \caption{(Left) Representation of the optical path of light in an isotropic GRIN medium, propagating in the direction of the wavefront defined by the wavevector angle $\theta$. (Right) Transformation of the physical plane and ray tracing by a conformal mapping $f$, recovering a medium gradient-index independent of $Y$-coordinate following the generalized Snell's law ${\mathfrak F=N\sin\theta=\mathrm{cst}}$. The angle $\theta$ is preserved through the transformation between the two spaces.}
    \label{fig:conform_scheme}
\end{figure*}

For the sake of simplicity, we illustrate the principle for an isotropic medium, and we write $n[\vec{x}(s)]$ its optical index. We also limit ourselves to a bi-dimensional system. Let ${z = x + {\rm i}y \in \mathbb{C}}$ with ${\vec{x}=(x, y)}$ be the spatial coordinates in the propagation plane, and $f$ be an analytic function (${f(z) = X + {\rm i}Y}$) that does not depend on the complex conjugate of $z$. This function thus defines a conformal map that locally preserves angles, transforming the ray parameter such that~\cite{nehari2012conformal}
\begin{equation} \label{eq:dS}
    {\rm d}S = {\rm d}s\left | \frac{{\rm d}f}{{\rm d}z} \right |,
\end{equation}
which corresponds to a transformed optical index~\cite{luneburg1966mathematical,leonhardt2006optical}
\begin{equation} \label{eq:N}
    N = n\left | \frac{{\rm d}f}{{\rm d}z} \right |^{-1}.
\end{equation}
Now, if there exists a particular function $f$ such that $N$ is independent of one of the coordinates, propagation in $f$-space is indistinguishable from that in a GRIN medium for which the optical index depends only on one coordinate. For instance, if ${N=N(X)}$, Fermat's invariant then takes the classical form of Snell's law for gradient-index media with 1d layered inhomogeneity. Using the notations from Fig.~\ref{fig:conform_scheme}, this writes~\cite{coleman2004generalization}
\begin{equation} \label{eq:FF}
    \mathfrak{F} = N(X)\frac{{\rm d}Y}{{\rm d}S}(X) = N(X) \sin\theta,
\end{equation}
which, using Eq.~\eqref{eq:N}, yields back in physical space
\begin{equation} \label{eq:fermat_fermat}
    \mathfrak{F} = \left | \frac{{\rm d}f}{{\rm d}z} \right |^{-1}n\sin\theta.
\end{equation}

\subsection{Symmetry in the velocity field}

Since, as stated above, our goal is to leverage the methods developed for GRIN media to model an effective rest-frame nonuniformity introduced by motion, the idea here is to find a system of coordinates in which the velocity field presents symmetries. 

\begin{figure}[b]
    \centering
    \includegraphics{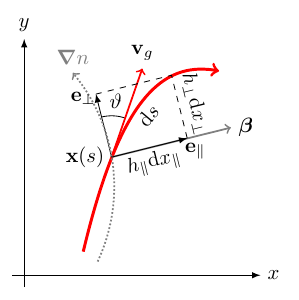}
    \caption{Sketch of the curvilinear parameterization based on the streamlines of the moving medium.}
    \label{fig:curve_scheme}
\end{figure}

Rather than attempting to solve this problem in a general fashion, we choose here to work in the orthogonal curvilinear coordinates constructed along the local direction of the body's motion. As depicted in Fig.~\ref{fig:curve_scheme}, we define 
${\big(O,\vec{e}_\perp\big[\vec x(s)\big],\vec{e}_\parallel\big[\vec x(s)\big] \big)}$ the local basis of this system, where ${\uvec{e}_\parallel = \vec\beta/|\vec\beta|}$ and 
${\vec{e}_\perp\perp\vec{e}_\parallel}$, and write
\begin{equation}
    h_\parallel \defeq \left|\frac{\partial\vec x}{\partial x_\parallel}\right| \quad \textrm{and} \quad h_\perp\defeq\left|\frac{\partial\vec x}{\partial x_\perp}\right|
\end{equation}
the associated scaling factors~\cite{pozrikidis2011introduction}. Although this particular system of coordinates based on streamlines does not generally guarantee a symmetry, it allows as we will show in the next section to recover with our model a number of results on the effect of motion known in canonical configurations. If one for instance assumes that the velocity profile does not depend on $x_\parallel$, that is ${\vec \beta=\beta(x_\perp) \uvec{e}_\parallel}$, which is notably true for circular flows, then Eqs.~\eqref{eq:FF}-\eqref{eq:fermat_fermat} give that the conserved quantity is
\begin{equation} \label{eq:frak}
    \mathfrak{F} = h_\parallel n \sin\theta.
\end{equation}

An important motivation for this choice of coordinates here is that it is particularly well suited to model drag. This is because the layers modeling our nonuniform velocity field are defined such that the velocity is tangent to the interface. Then, since by definition the ray parameter writes in this coordinates system
\begin{equation} \label{eq:ds}
    {\rm d}s^2 = h_\perp^2{\rm d}x_\perp^2+h_\parallel^2{\rm d}x_\parallel^2,
\end{equation}
we get that the inclination of the ray with respect to the interface normal, that is $\vartheta$, simply writes
\begin{equation} \label{eq:vartheta}
    \tan \vartheta(s) = \frac{h_\parallel}{h_\perp} \frac{{\rm d}x_\parallel}{{\rm d}x_\perp}(s).
\end{equation}

Note that it is still possible within this method to combine the effect of velocity and that of material inhomogeneities. The one constraint, however, is that the gradient characterizing material inhomogeneities in the medium's rest-frame has to be normal to the velocity at all points, so as to preserve the symmetry in streamline coordinates. This is for instance true of a circular flow with a radial density gradient.

\subsection{Light path for symmetrical velocity fields}

Going back to our goal to model the light path,  the invariant associated with the velocity symmetry Eq.~\eqref{eq:frak} can be used to eliminate the $n\sin\theta$ dependencies in Eqs.~\eqref{eq:evaluated1}-\eqref{eq:evaluated2}. Eq.~\eqref{eq:vartheta} is then used together with Eq.~\eqref{eq:nonuniform_Psi} to get
\begin{subequations} \label{eq:curv_eq}
\begin{equation} \label{eq:curv_eq_anisotrop}
    \frac{{\rm d}x_\parallel}{{\rm d}x_\perp} =  \frac{\gamma h_\perp}{h_\parallel}\Bigg\{c\beta \frac{\partial \mathcal K'}{\partial \omega'}(\omega', k_{\parallel}') - \frac{\partial \mathcal K'}{\partial k_{\parallel}'}(\omega', k_{\parallel}')\Bigg\}
\end{equation}
with
\begin{align}
    \omega' &= \big(1-h_\parallel^{-1}\mathfrak F\beta \big)\gamma \omega, \\ 
    k_{\parallel}'&=\big(h_\parallel^{-1}\mathfrak F-\beta \big)\gamma \omega/c.  
\end{align}              
\end{subequations}
Here $h_\perp$, $h_\parallel$, $\beta$ and $\gamma$ are in general functions of $x_\perp$ and $x_\parallel$. Eqs.~\eqref{eq:curv_eq} conveniently provides an equation for the evolution of a light ray in a nonuniformly moving anisotropic media, without the need to compute explicitly the wavevector at each point. One instead only needs the be able to determine the invariant Eq.~\eqref{eq:frak} at a certain point along the ray.

\section{Effect of linear and rotational motions on light in isotropic media} \label{sec:app}

Before exposing in Sec.~\ref{sec:plasma} the new capabilities brought by the approach proposed in Sec.~\ref{sec:sym} to model the effect of a nonuniform motion on the light path in anisotropic media, we first demonstrate in this section how it recovers known results from the literature in the limit of a medium that is isotropic in its rest-frame. Since any motion can be decomposed into the sum of linear and rotational motions, and because the velocity fields associated with these canonical motions present symmetries, we consider here these two particular motions as examples. To shorten equations we introduce the notation ${[f(\xi)]_a^b=f(b)-f(a)}$.

For a medium that is isotropic in the comoving frame Eq.~\eqref{eq:N_prime} can be rewritten as 
\begin{equation} \label{eq:K_prime_iso}
    \mathcal K'\left(\omega',k_\parallel'\right)=\sqrt{\left(\frac{\omega' n'(\omega')}{c}\right)^2-k_\parallel'^2}.
\end{equation}
In this limit the general drag formula Eq.~\eqref{eq:nonuniform_Psi} reduces to the simpler form
\begin{equation} \label{eq:non_uniform_drag_angle_isotrope}
    \tan \vartheta(s) = \gamma \frac{n'(\omega')n_g'(\omega') \beta\left(1-\beta n \sin\theta\right )+n \sin \theta-\beta}{\sqrt{n'^2(\omega')\left(1-\beta n \sin\theta\right )^2-\left(n \sin\theta-\beta\right )^2}}
\end{equation}
where we have introduced ${n_g'=n'+\omega'{\rm d}n'/{\rm d}\omega'}$ the rest-frame group index evaluated at ${\omega'=\gamma\omega(1 - \beta n \sin\theta)}$, and where all quantities are functions of $\vec{x}(s)$. This in turn leads to the result
\begin{equation} \label{eq:curv_eq_isotrop}
    \frac{{\rm d}x_\parallel}{{\rm d}x_\perp} = \frac{\gamma h_\perp}{h_\parallel}\frac{n'(\omega')n_g'(\omega') \beta\big(h_\parallel-\mathfrak{F}\beta \big )-h_\parallel\beta+\mathfrak{F}}{\sqrt{n'^2(\omega')\big(h_\parallel-\mathfrak{F}\beta \big)^2-\big(\mathfrak{F}-h_\parallel\beta\big )^2}},
\end{equation}
which is the isotropic version of Eq.~\eqref{eq:curv_eq_anisotrop}.

\subsection{Nonuniform linear motion}

Consider as a first example a nonuniform linear motion described by ${\vec{\beta} = \beta(x) \uvec{e}_y}$ in a Cartesian coordinate system $(x,y)$. This is the differentially sheared configuration studied by Lerche~\cite{lerche1974passage}. In the GRIN representation adopted here, the motion is the source of an effective inhomogeneity along $x$. The velocity aligned curvilinear coordinates defined in the previous section are here simply ${(x_\perp,x_\parallel)=(x,y)}$. The scale factors associated with the conformal identity ${f(z)=z}$ are ${h_\parallel=h_\perp=1}$.

From Eq.~\eqref{eq:frak}, Fermat's invariant simply writes in this case ${\mathfrak F_x=n\sin\theta}$, which is the differential form of the classical Snell's law~\cite{trager2012springer}. From Eq.~\eqref{eq:curv_eq_isotrop} the equation for the optical path $y(x)$ writes
\begin{subequations} \label{eq:4556}
\begin{equation} 
    \frac{{\rm d}y}{{\rm d}x} = \gamma(x)\frac{n'(\omega') n_{g}'(\omega') \beta(x)\left[1-\mathfrak{F}_x\beta(x)\right]-\beta(x)+\mathfrak{F}_x}{\sqrt{n'^2(\omega')\left[1-\mathfrak{F}_x\beta(x)\right ]^2-\left[\beta(x)-\mathfrak{F}_x\right ]^2}}
\end{equation}
where
\begin{equation} \label{eq:44665}
    \omega'= [1-\mathfrak{F}_x\beta(x)]\gamma(x)\omega.
\end{equation}
\end{subequations}
This result is found to be consistent with the results obtained by Lerche~\cite{lerche1974passage}. We also note that Eq.~\eqref{eq:4556} bears resemblance with results previously obtained for acoustic rays in linearly-moving inhomogeneous media (see e.g. Ref.~\cite{hohenwarter2000snell}), but differs due notably to the use here of Lorentz transformation rather than the Galilean transformation used for sound waves.

Eq.~\eqref{eq:4556} is also consistent with results known for a uniform motion. Specifically, $\beta$ and $\gamma$ are in this case constants, so that from Eq.~\eqref{eq:4556} ${{{\rm d}y}/{{\rm d}x} = \mathrm{cst}}$. This is consistent with the straight path expected for the beam deflection predicted for the Fresnel drag for a uniform linear motion. For refraction at an interface, Fermat's invariant can be written in terms of the incident beam parameters. Precisely, from Eq.~\eqref{eq:generalized_snell2}, one gets ${\mathfrak F_x=\bar n_i\sin\theta_i}$. Eq.~\eqref{eq:4556} thus becomes the aberration formula of Ko and Chuang~\cite{ko1978passage} or, by setting ${\mathfrak F_x=0}$, its simplified form for normal incidence as historically derived by Player~\cite{player1975}.

\begin{figure}
    \centering
    \includegraphics{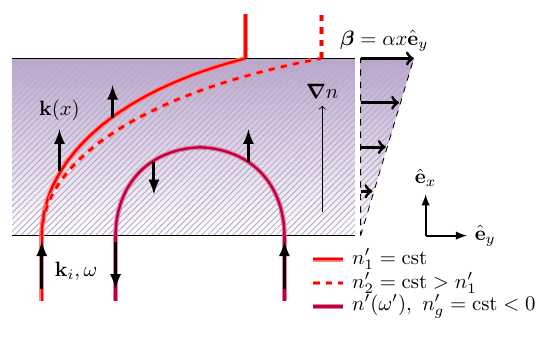}
    \caption{Light path in an isotropic dielectric in linear motion ${\vec\beta = \alpha x \uvec{e}_y}$, for an incident ray at normal incidence. The solid and dashed red rays represent the path for two nondispersive media with different constant optical indices, highlighting the tendency of light to avoid high-index regions. The purple ray shows the path in a dispersive medium with anomalous dispersion. Since the transverse component of the wavevector is conserved, it remains, in this particular case of normal incidence, perpendicular to the velocity all along the path.  Just like for conventional material-GRIN turning point, the wavevector flips sign at the turning point.}
    \label{fig:linear_drag}
\end{figure}

As a particular example of the application of the ray tracing equation Eq.~\eqref{eq:4556}, let us consider a dielectric medium moving with a linear velocity profile ${\beta(x)=\alpha x}$ with ${\alpha\in\mathbb R}$. We write $y(0)=y_i$ and focus as illustrated in Fig.~\ref{fig:linear_drag} on the particular case of normal incidence, for which ${\mathfrak F_x=0}$. Integrating Eq.~\eqref{eq:4556} yields directly
\begin{equation} 
    y(x) = y_i+\int_0^x \frac{(n_g'-1/n')\alpha\xi}{\sqrt{\left[ 1-(\alpha\xi)^2\right ]\left[ 1-(\alpha\xi/n')^2\right ]}}~\textrm{d}\xi. \label{eq:989}
\end{equation}
In the limit of a nondispersive medium (${n_g'=n'=\mathrm{cst}}$), one finds
\begin{equation} 
    y(x) =y_i+\left(n'-\frac{1}{n'}\right)\frac{n'}{\alpha}\left[\mathrm{arcsinh}\sqrt{\frac{1-\xi^2}{ n'^2-1}}\right]_{\alpha x}^0. \label{eq:990}
\end{equation}
This can be rewritten to leading order in $\beta$---that is for nonrelativistic motion---as
\begin{equation}
    y(x) = y_i+\frac{\alpha}{2}\left(n'-\frac{1}{n'}\right) x^2.
\end{equation}

This response of light to a linear velocity profile is illustrated in Fig.~\ref{fig:linear_drag}, where  the solution Eq.~\eqref{eq:990} is plotted for two different rest-frame refractive indexes $n'$. Note interestingly that even though the path is deviated, the wavefront remains parallel to the motion, as already pointed out by Player~\cite{player1975}. This comes from the fact that Fermat's invariant is null at normal incidence. Trying to make sense of the effect of the velocity, it can easily be shown that the lab-frame wave index $n$ for a ray at normal incidence on this simple isotropic media and the velocity field considered writes~\cite{langlois2025fresnel}
\begin{equation}
    n(x) = \sqrt{\frac{n'^2-\beta^2(x)}{1-\beta^2(x)}},
\end{equation}
which one verifies is an increasing function of $x$. We thus find here that Fermat's principle Eq.~\eqref{eq:fermat_principle}, which guarantees that rays avoid regions of high refractive index so as to minimize the optical path, carries over to the effective index associated with velocity. 

Similarly to GRIN media theory, possible poles in the integrand of the optical path equation correspond to turning points along the trajectory. This behavior is associated with a decrease of the wavevector, which eventually becomes zero and then flips sign. A rapid analysis of Eq.~\eqref{eq:989} shows that there cannot be singularity for a constant rest-frame optical index. Indeed, poles write ${\beta(x)=1}$ or ${\beta(x)=\bar n}$, neither of which is possible for a nondispersive medium since in this case ${\bar n>1}$~\cite{ko1978passage}. On the other hand frequency dispersion, in allowing for ${n'(\omega')<1}$, introduces the possibility for turning points. An example is the ideal model for a dielectric under electromagnetically induced transparency (EIT) conditions~\cite{fleischhauer2005electromagnetically} suggested by Leonhardt and Piwnicki~\cite{leonhardt2000relativistic,leonhardt2001slow}. Following their work, we similarly write the rest-frame dielectric response in the immediate vicinity of a given characteristic frequency $\omega_0$ as
\begin{equation} \label{eq:p45}
    n'(\omega')\sim\sqrt{1+2n_g'\left(\frac{\omega'}{\omega_0}-1\right)}
\end{equation}
where $n_g'$ with ${|n_g'|\gg1}$ represents the dispersionless rest-frame group index for ${\omega=\omega_0}$. Since at normal incidence ${\omega'=\gamma(x)\omega}$, the term in parenthesis under the square root in Eq.~\eqref{eq:p45} is positive and increases with $x$ if considering a probe beam ${\omega\sim\omega_0^+}$ in the lab-frame. Then, in the particular case of anomalous dispersion ${n_g'<0}$, the optical index Eq.~\eqref{eq:p45} becomes a decreasing function of $x$, which varies from $1$ for ${x=0}$ to $0$ for a critical $x$ (i.e.,~$\beta$). There thus exists in between a point $x_c$ where ${\beta(x_c)=n'(x_c)}$. From Eq.~\eqref{eq:989} this point corresponds to a turning point where the beam's propagation direction reverses. This is the behavior illustrated by the half-circle ray in Fig.~\ref{fig:linear_drag}. Note that in this specific case deflection is \textit{anomalous}~\cite{banerjee2022anomalous}, i.e., the drag is along the direction opposite to the medium’s movement. Such turning point is a relativistic analog to a material turning point, although here the beam does not emerge at the same location. We note that this interesting property could in principle be used as a tool to diagnose the velocity field, possibly probing different depths and velocities through a frequency scan.

\subsection{Rotation}

A circular rotation around a given axis, chosen here to be $\uvec{e}_z$, is simply parameterized in the polar orthonormal basis ${(O,\uvec{e}_r,\uvec{e}_\varphi)}$ by
\begin{equation} \label{eq:rotation}
    \vec \beta=\frac{r\Omega(r)}{c} \uvec{e}_\varphi,
\end{equation}
where we have introduced the angular velocity $\Omega(r)$. The associated conformal map is $f(z)=\ln(z)$~\cite{coleman2004generalization}. The velocity aligned curvilinear coordinates write ${(x_\perp,x_\parallel)=(r,\varphi)}$ with the associated scale factors ${h_\perp=1}$ and ${h_\parallel=r}$.  From Eq.~\eqref{eq:frak} one then gets
\begin{equation} \label{eq:bouguer}
    \mathfrak F_r=r n \sin\theta.
\end{equation}
This relation, sometimes known as Bouguer's law, is Fermat's invariant for radial GRIN media. It is the optical analogue to the conservation of angular momentum for a particle in a central force field~\cite{born2013principles}. Plugging Eq.~\eqref{eq:bouguer} into Eq.~\eqref{eq:curv_eq_isotrop} gives the optical path equation
\begin{subequations} \label{eq:mpa}
\begin{equation}\label{eq:mpa1}
    \frac{\textrm d\varphi}{\textrm d r} =  \frac{\gamma(r)}{r}\frac{n'(\omega')n_g'(\omega') \beta(r)\big[r-\mathfrak{F}_r\beta(r) \big ]-r\beta(r)+\mathfrak{F}_r}{\sqrt{n'^2(\omega')\big[r-\mathfrak{F}_r\beta(r) \big]^2-\big[\mathfrak{F}_r-r\beta(r)\big ]^2}}
\end{equation}
with
\begin{equation}
    \omega' =  \big[r-\mathfrak F_r\beta(r) \big]\frac{\gamma(r)}{r}\omega.\label{eq:omegap_rot}
\end{equation}
\end{subequations}

In the limit ${\beta=0}$ (and therefore ${n'\to\bar n}$) one finds 
\begin{equation} \label{eq:radial_GRIN}
    \frac{\textrm d\varphi}{\textrm d r} =  \frac{\mathfrak{F}_r}{r\sqrt{\bar n^2(r)r^2-\mathfrak{F}_r^2}},
\end{equation}
which we recognize as the ray equation known for a static radial GRIN medium $\bar{n}(r)$~\cite{born2013principles}. In the remaining of this section we analyze the effect of rotation for different angular velocity profiles. 

\subsubsection{Rigid-body rotation}

The literature on the effect of rotation on waves in rotating dielectrics generally assumes a rigid body motion. This assumption is consistent with the fact that experimental studies classically used solid glass rods, as in Jones' Fresnel drag~\cite{jones1972fresnel,jones1975} and polarization drag~\cite{jones1976rotary} experiments. In this particular case one simply takes ${\Omega(r)=\mathrm{cst}=\Omega_0}$ in Eq.~\eqref{eq:rotation}. 

\begin{figure}
    \centering
    \includegraphics{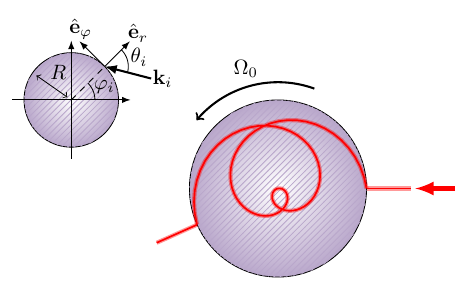}
    \caption{(Left) Parametrization of a rotating disk for any incidence. (Right) Illustration of the Archimedean optical path followed by light in a highly-dispersive and/or fast-rotating dielectric undergoing rigid-body rotation.}
    \label{fig:rotation}
\end{figure}

Consider as shown in Fig.~\ref{fig:rotation} the case of a beam at normal incidence on the rotating medium. In this particular configuration one finds from Eq.~\eqref{eq:bouguer} $\mathfrak F_r=0$, and in turn from Eq.~\eqref{eq:omegap_rot} $\omega'=\gamma(r)\omega$. The light path in a dielectric disk of radius $R$ then writes
\begin{equation}
    \varphi(r) = \varphi_i+\int_r^R \frac{\Omega_0}{c}\frac{(n_g'-1/n')}{\sqrt{\left[1-\left({\Omega_0 \xi}/c\right )^2\right]\left[1-\left( {\Omega_0 \xi}/{cn'}\right )^2\right]}}\textrm{d}\xi. \label{eq:r1}
\end{equation}

In the limit of a nondispersive medium ${n_g'=n'=\mathrm{cst}}$, the solution writes
\begin{equation}
    \varphi(r) = \varphi_i + \left(n'-\frac 1{n'}\right)\Big[F\left(\arcsin\left(\Omega_0 \xi/c \right ),1/n'^{2} \right )\Big]_r^R \label{eq:r2}
\end{equation}
where $F$ is the Jacobi incomplete elliptic integral of the first kind~\cite{abramowitz1968handbook} and ${\varphi_i=\varphi(R)}$. As illustrated in Fig.~\ref{fig:rotation}, Eq.~\eqref{eq:r2} is the equation of an Archimedean spiral. {Just like there was no pole for the ray trajectory in a nondispersive medium in linear motion, there is no pole in Eq.~\eqref{eq:mpa1} at normal incidence as $\beta<1<n'$. A consequence is that there is no turning point, which implies as seen in Fig.~\ref{fig:rotation} that the radial position of the ray reaches $r=0$ before moving radially outward.} 

For a dispersive medium Eq.~\eqref{eq:r1} yields to first order in $\beta$
\begin{equation}
    \varphi(r) = \varphi_i + \frac{R\Omega_0}{c} \left(n_g'(\omega)-\frac 1{n'(\omega)}\right)\left(1-\frac rR\right).
\end{equation}
We recognize in this result the Fresnel coefficient ${(n_g'-1/n')}$, with here both $n_g'$ and $n'$ functions of $\omega$. This result is the analogue for a rotation to the transverse drag theorized by Player~\cite{player1975} and observed by Jones~\cite{jones1972fresnel,jones1975} for a uniform linear motion. 

\subsubsection{Vortex rotation}

Another classical rotating configuration is an irrotational vortex flow, which is characterized by ${\beta(r)\propto r^{-1}}$, or equivalently by ${\Omega(r)\propto r^{-2}}$~\cite{lamb1945hydrodynamics}. Such profile, however, presents a singularity on axis. To avoid this issue, Leonhardt and Piwnicki suggested a 'relativistic' profile by adding a Lorentz factor $\gamma$ to the velocity. The velocity field of this relativistic vortex accordingly writes~\cite{leonhardt1999optics}
\begin{equation}
    \boldsymbol \beta(r) = \frac{1}{\gamma(r)}\frac{\Gamma}{2\pi cr}\uvec{e}_\varphi 
\end{equation}
with
\begin{equation}
\gamma(r) = \sqrt{1+\frac{\Gamma^2}{4\pi^2c^2r^2}}.
\end{equation}
Here ${\Gamma\in\mathbb R}$ is the vortex circulation, giving a constant angular momentum per unit mass ${cr\gamma(r)\beta(r)=\Gamma/2\pi}$ relative to the rotation axis. 

\paragraph{Determination of Fermat's invariant}

\begin{figure*}
    \centering
    \includegraphics{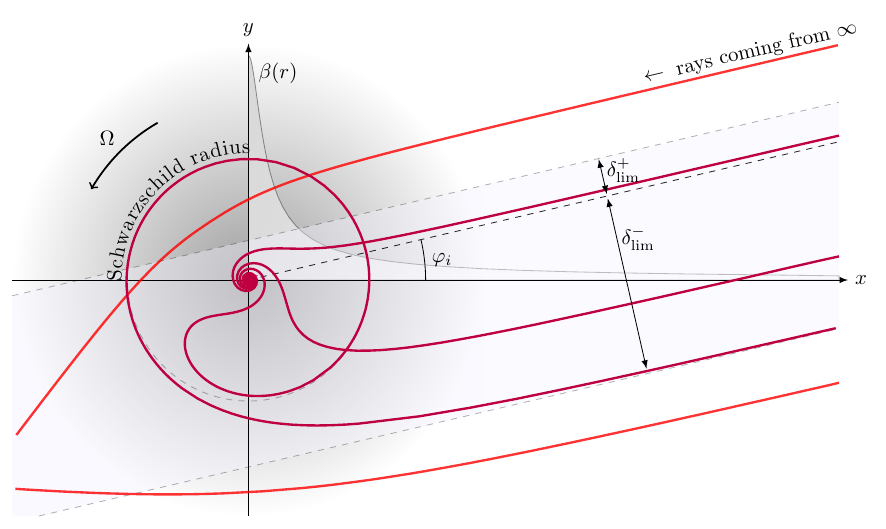}
    \caption{Optical path in a relativistic nondispersive vortex of rays incident from infinity with different impact parameters relative to an incidence axis $(O,\varphi_i)$ (shown as the dashed black line). The limiting values $\delta_\mathrm{lim}^+$ and $\delta_\mathrm{lim}^-$ define the region within which rays are not simply deflected (red beams) but instead captured by the vortex (purple beams), akin to a black hole. Rays close to $\delta_\mathrm{lim}^-$ undergo significant deflection along the Schwarzschild radius before being trapped by the vortex.}
    \label{fig:vortex}
\end{figure*}

Consistent with Leonhardt and Piwnicki's work~\cite{leonhardt1999optics}, we consider a ray incident from infinity, where it is assumed to travel in a straight line since ${\beta(r\to\infty)=0}$. A downside of this hypothesis in our model is that Fermat's invariant from Eq.~\eqref{eq:bouguer} is \textit{a priori} unknown since neither the effective index nor the wavevector are known. The symmetry of the velocity field does however guarantee an invariant, and we thus simply write for now the evolution of the angular position as a function of this invariant $\mathfrak F_r$. Specifically, by integration of Eq.~\eqref{eq:mpa}, it comes
\begin{multline} \label{eq:123}
    \varphi_\mathrm{in}(r) = \varphi_i\\+\int_r^\infty\frac{\gamma(\xi)}{\xi}\frac{ n'^2 \beta(\xi)\left[\xi-\mathfrak F_r\beta(\xi)\right]+\mathfrak F_r-\xi\beta(\xi)}{\sqrt{n'^2\left[\xi-\mathfrak F_r\beta(\xi)\right]^2-\left[\mathfrak F_r-\xi\beta(\xi)\right]^2}}\textrm{d}\xi.
\end{multline}
where, as depicted in Fig.~\ref{fig:vortex}, we write $\varphi_i$ the inclination of the incident beam at infinity. The subscript '$\mathrm{in}$' indicates that Eq.~\eqref{eq:123} represents the angular position of the ray as it approaches the center of rotation. Once the point of closest approach characterized by the polar coordinates $(\varrho,\varphi_\varrho)$ has been reached, the angular position of the ray simply writes
\begin{equation} \label{eq:124}
    \varphi_\mathrm{out}(r) = 2\varphi_\varrho - \varphi_\mathrm{in}(r).
\end{equation}
Naturally the point of closest approach is reached when the denominator of Eq.~\eqref{eq:123} vanishes, i.e.
\begin{equation} \label{eq:159}
    n'^2\left[\varrho-\mathfrak F_r\beta(\varrho)\right]^2-\left[\mathfrak F_r-\varrho\beta(\varrho)\right]^2=0.
\end{equation}

If we now consider, similarly again to Leonhardt and Piwnicki~\cite{leonhardt1999optics}, the impact parameter of this ray which we write here $\delta$, the angular position must verify the condition 
\begin{equation} \label{eq:753}
    \lim_{r\to\infty} r \sin\left[\varphi(r)-\varphi_i\right] = \delta,
\end{equation}
which allows determining Fermat's invariant. Practically, Taylor expanding the integrand of Eq.~\eqref{eq:123} for large $r$ leads to
\begin{equation}
    \varphi(r)-\varphi_i\sim\left[\frac{\mathfrak F_r}{n'}+\left(n'-\frac{1}{n'}\right)\frac{\Gamma}{2\pi c} \right ] \int_r^\infty \frac{{\rm d}\xi}{\xi^2}
\end{equation}
which upon integration gives
\begin{equation}
    r\sin\left[\varphi(r)-\varphi_i\right]\sim\frac{\mathfrak F_r}{n'}+\left(n'-\frac{1}{n'}\right)\frac{\Gamma}{2\pi c}.
\end{equation}
Identification with Eq.~\eqref{eq:753} then immediately yields
\begin{equation}
\label{Eq:invariant_Fr}
    \mathfrak F_r=n'\delta-\left(n'^2-1\right)\frac{\Gamma}{2\pi c}.
\end{equation}
Compared to the 'naive' expression $n'\delta$ that one may infer if neglecting the misalignment between the lab-frame wavevector and the group velocity, and considering that the effective optical index $n$ reaches the rest-frame index $n'$ far off-axis, the Fermat invariant is found here to be shifted by ${(n'^2-1)\Gamma/(2\pi c)}$. This deviation was identified by Leonhardt and Piwnicki as a manifestation of the optical Aharonov-Bohm effect~\cite{leonhardt1999optics}, noting that $n'\delta$ here plays the role of their modified angular momenta $l_\textrm{AB}$.

\paragraph{Comparison with known results}

We can now plug the result for the invariant Eq.~\eqref{Eq:invariant_Fr} into Eqs.~\eqref{eq:123} and~\eqref{eq:124} to obtain expressions for the ray polar coordinates as a function of $\delta$ and $\Gamma$, and see how our results match Leonhardt and Piwnicki's ray tracing solutions~\cite{leonhardt1999optics}.

An important finding of Leonhardt and Piwnicki is the result that small enough impact factors lead to ray paths that converge to the axis. This has led relativistic vortices to be coined 'optical black holes'. Such rays are by nature characterized by the absence of a point of closest approach (or turning point). Rewriting Eq.~\eqref{eq:159} as an equation for the impact parameter as a function of this point of closest approach, we find 
\begin{equation} \label{eq:987}
    \delta_\pm = \left(n'-\frac{1}{n'} \right )\frac{\Gamma}{2\pi c}+ \frac{\varrho}{n'}\frac{n'\pm\beta(\varrho)}{n'\beta(\varrho)\pm1}
\end{equation}
where one shows that the $+$ and $-$ solutions correspond respectively to positive and negative impact parameters, assuming a positive rotation relative to the chosen coordinate system $\Gamma>0$. Impact parameters leading to rays converging on axis thus correspond to those for which Eq.~\eqref{eq:987} has no solution. 

Analyzing the solutions of Eq.~\eqref{eq:987}, one finds that the positive branch $\delta_+$ reaches its minimum for ${\rho=0}$, that is ray with positive impact parameters can asymptotically approach the rotation axis, while still emerging out and back to infinity. For ${\rho=0}$  Eq.~\eqref{eq:987} immediately gives a limit positive impact parameter
\begin{equation}
    \delta_\mathrm{lim}^+=\left(n'-\frac{1}{n'} \right )\frac{\Gamma}{2\pi c}.
\end{equation}
In contrast, the negative branch $\delta_-$ has no solution for sufficiently low $\rho$, since $\delta_-(\rho)$ exhibits a maximum. The limit negative impact parameter is accordingly $\delta_\mathrm{lim}^-=\delta_-(\varrho_s)$ with $\varrho_s$ defined by
\begin{equation} \label{eq:condi}
    \frac{{\rm d}\delta_-}{{\rm d}\varrho}(\varrho_s)\defeq 0.
\end{equation}
Because by definition there are no ray paths leaving the vortex for negative impact parameters larger than $\delta_\mathrm{lim}^-$, the radius $\varrho_s$ is referred to as the 'optical Schwarzschild radius'. Computing the derivative from Eq.~\eqref{eq:987}, one finds after lengthy but straightforward calculations that 
\begin{subequations} \label{eq:0001}
\begin{equation}
    \varrho_s = \frac{\Gamma}{2\pi c}\sqrt{\frac 1{a^2}-1}
\end{equation}
where
\begin{align}
    a&=\frac{n'}{3(n'^2-1)}\left[2\sqrt{6n'^2-5}\sin\left(\frac 13\arcsin b \right ) -1\right ], \\
    b&=\frac{45n'^4-70n'^2+27}{2n'^2(6n'^2-5)^{3/2}},
\end{align}
\end{subequations} 
or, in the limit of large $n'$,
\begin{equation} \label{eq:0002}
    \varrho_s \sim \frac{\Gamma n'}{\pi c}.
\end{equation}
{Noting the trigonometric identity $\sin(\arcsin(x)/3)=\cos([4\pi+\arccos(-x)]/3)$,} Eqs.~\eqref{eq:0001} and~\eqref{eq:0002} are precisely the results obtained by Leonhardt and Piwnicki~\cite{leonhardt1999optics}. This demonstrates the consistency of the two methods, although as already mentioned Leonhardt and Piwnicki themselves tackled the problem differently by using the analogy between a moving dielectric and a gravitational field captured in Gordon's metric~\cite{gordon1923lichtfortpflanzung}. 

Putting these pieces together, one finds that the limit negative impact parameter 
\begin{equation} \label{eq:7533}
    \delta_\mathrm{lim}^-=\left(n'-\frac{1}{n'} \right )\frac{\Gamma}{2\pi c}+\frac{\varrho_s}{n'}\frac{n'-\beta(\varrho_s)}{n'\beta(\varrho_s)-1}
\end{equation}
differs from its positive counterpart $\delta_\mathrm{lim}^+$. This symmetry breaking is the signature of rotation, exposing an intrinsic difference between black hole and effective curvatures.

To conclude this section, we verify in a similar manner that the present work recovers the results subsequently obtained by Leonhardt and Piwnicki when extending their nondispersive theory to model the particular case of dispersive media verifying Eq.~\eqref{eq:p45}~\cite{leonhardt2000relativistic,leonhardt2001slow}. The theory proposed here, however, captures indifferently other and more complex forms of optical index $n'(\omega')$ for a dispersive dielectric.

\section{Application to nonuniformly moving magnetized plasmas} \label{sec:plasma}

To conclude this study and illustrate the additional capabilities of the theory proposed in this work, we finally consider the effect of motion in a dispersive anisotropic medium, taken here to be a magnetized plasma. We stress here that this is only meant to be a first illustration of the capabilities and results of the proposed model, and not an attempt at quantifying drag effects in moving plasmas for realistic conditions. Such a study is left for future work.

\subsection{Rotating magnetized plasma column}

To make things simpler, we first consider the particular case of modes for which the rest-frame optical index does not depend on the wavevector ${n'(\omega',\uvec{k}')\to n'(\omega')}$. In this case the phase and group velocities in the comoving-frame are collinear, and, as demonstrated earlier, Eq.~\eqref{eq:nonuniform_Psi} reduces to the simpler form Eq.~\eqref{eq:non_uniform_drag_angle_isotrope}. For that we consider a quasi-neutral magnetized plasma column of radius $R$ rotating at constant angular velocity $\Omega_0$ where the external static magnetic field $B_0$ is assumed to be uniform and parallel to the column's axis, and a wave incident from the surrounding vacuum region with wavevector in the plane perpendicular to the axis, that is ${\vec{k} \perp \vec{B}_0}$. This way the excited modes are classically the ordinary (O) and extraordinary (X) modes of a magnetized plasma~\cite{stix1992waves}, guaranteeing that the phase and group velocities in the comoving-frame are indeed collinear.

Neglecting classically any manifestation of motion in the instantaneous rest-frame~\cite{van1976electromagnetic,player1976dragging,leonhardt1999optics,gotte2007dragging,Gueroult2019}---that is, with our notation, assuming ${n'=\bar{n}}$---, the refractive indices in $\Sigma'$ of these two modes write~\cite{stix1992waves,krall1973principles}
\begin{equation}\label{eq:index_0}
    n_O'(\omega')=\sqrt{1-\left(\frac{\omega_p}{\omega'}\right)^2}
\end{equation}
and
\begin{subequations} \label{eq:xmode}
\begin{equation}\label{eq:index_X}
    n_X'(\omega')=\sqrt{\frac{(\omega'^2-\omega_L^2)(\omega'^2-\omega_R^2)}{(\omega'^2-\omega_{UH}^2)(\omega'^2-\omega_{LH}^2)}}.
\end{equation}
Here 
\begin{equation}
    \omega_{R/L} = \mp \frac{\Omega_{ce}+\Omega_{ci}}{2} + \frac{1}{2}\sqrt{(\Omega_{ce}-\Omega_{ci})^2+4\omega_{p}^2}
\end{equation}
are the right- and left-cutoffs while
\begin{equation}
    \omega_{UH/LH} = \left[\frac{\varpi_e^2+\varpi_i^2}{2}\pm \frac{1}{2}\sqrt{(\varpi_e^2-\varpi_i^2)^2+4\omega_{pe}^2\omega_{pi}^2}\right]^{1/2}
\end{equation}
\end{subequations}
are the upper- and lower-hybrid frequencies, and we have defined $\omega_{p}^2=\sum_s \omega_{ps}^2$ and $\varpi_s^2 = \omega_{ps}^2+\Omega_{cs}^2$
with 
\begin{equation}
    \omega_{ps}(r)=\sqrt{\frac{N(r)q_s^2}{\epsilon_0 m_s}} \quad \textrm{and} \quad \Omega_{cs}=\frac{q_s B_0}{m_s}
\end{equation}
respectively the plasma and cyclotron frequency of each plasma species $s$ with particle's mass $m_s$ and charge $q_s$. Note here that these frequencies, together with the mass, the density, and the magnetic field that appear in it, should technically all be written in the rest-frame, and thus be primed quantities. Yet, those differ from unprimed quantities by at most a factor $\gamma$~\cite{chawla1966note}. Because we focus here on the nonrelativistic limit we omit these differences and use for simplicity unprimed quantities.

Note finally here that these modes are dispersive, beyond what may be captured by Leonhardt and Piwnicki's model~\cite{leonhardt2000relativistic,leonhardt2001slow}. 

\subsubsection{High-frequency wave drag from dispersion}

\begin{figure}
    \centering
    \includegraphics{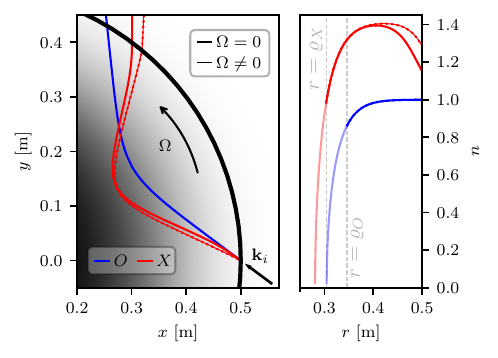}
    \caption{Ray trajectories for a wave incident on a rotating plasma cylinder (left) and corresponding lab-frame wave index as a function of radius (right), for a magnetic field ${B_0=0.1}$ T and plasma parameters $(N_0,\sigma,\Omega_0)=(10^{19}~\mathrm{m}^{-3}, 0.1~\mathrm{m}, 3~10^5~\mathrm{rad.s}^{-1})$. The blue curve correspond to the O mode, while the red curve corresponds to the X mode. Square markers on thin lines are absent rotation, whereas thick solid lines account for rotation. The color scale represents the radial density gradient. The incident wavevector is here $(k_i, \theta_i)=$ ($58.62$ m$^{-1}$, $0.64$ rad), which corresponds to a wave frequency just below the electron cyclotron frequency.}
    \label{Fig:OX}
\end{figure}

Let us first illustrate the effect of motion associated with dispersion on high-frequency waves. For this we consider an inhomogeneous plasma with \begin{equation}
    N(r)=N_0\exp\left(-\frac{r^2}{2\sigma^2} \right)
\end{equation}
where $N_0$ is the density on axis (maximum) and $\sigma$ characterized
the steepness of the profile, and assume that the plasma is slightly overdense on axis, i.e., ${|\Omega_{ce}|\simeq 2\omega_{pe}(r=0)}$, but with a rapid density fall off ${\sigma\leq R}$ so that ${|\Omega_{ce}|\gg\omega_{pe}(R)}$ at the edge. To expose dispersion effect we consider specifically a wave with frequency $\omega$ just below the electron cyclotron frequency. Indeed for a largely underdense edge plasma ${\omega_{UH}\sim|\Omega_{ce}|}$ and ${|{\rm d} n_X'/{\rm d}\omega'|}$ is large. The trajectories of O and X modes in these conditions, obtained by integrating Eq.~\eqref{eq:mpa}, are plotted in Fig.~\ref{Fig:OX}. We stress here that this is a toy problem meant to illustrate an effect, rather than an attempt at modeling a realistic configuration. It is notably acknowledged that choosing a wave frequency close to the resonance, even if only locally at the edge, poses a number of modeling challenges. It is only done here to emphasize the effect of dispersion.

Starting with the trajectories without rotation (dashed curves in Fig.~\ref{Fig:OX}), we see that both the O and X modes are reflected. This behavior can be interpreted in terms of the wave indexes. For the O mode the negative radial density gradient leads to a wave index Eq.~\eqref{eq:index_0} that increases monotonically with $r$, as shown in the right panel of Fig.~\ref{Fig:OX}. The O ray is accordingly progressively turned around. The evolution of the X mode index Eq.~\eqref{eq:index_X} is more intricate. Because we chose $\omega$ just below $|\Omega_{ce}|$, the wave frequency is close to the upper-hybrid resonance at the low density plasma edge. In this case, one can show that 
the wave index $n_X'(\omega)$ first increases with ${\omega_{pe}\ll|\Omega_{ce}|}$. As a consequence, the ray experiences an increasing wave index as it enters the plasma and encounters denser layers. The ray trajectory accordingly first becomes more and more radial. As ${\omega_{pe}\ll|\Omega_{ce}|}$ becomes larger though, the upper-hybrid frequency progressively moves away from the wave fixed frequency, leading to a subsequent decrease of $n_X'(\omega)$. This behavior is illustrated in the right panel of Fig.~\ref{Fig:OX}. Past this point the behavior is analogous of that of the O mode, with the ray being progressively turned around and reflected by the density gradient.  

Moving on to the effect of rotation, we first confirm in Fig.~\ref{Fig:OX} that rotation has no effect on the O mode. This result was expected since the dispersion relation of the O mode is Lorentz invariant~\cite{ko1978passage,langlois2025fresnel,braud2025ray}. Fig.~\ref{Fig:OX} however reveals a modification of the X mode ray trajectory. Analyzing this behavior, we note that rest-frame frequency $\omega'$ is now upshifted or downshifted as a result of motion. For the particular case considered in Fig.~\ref{Fig:OX}, ${\vec{k}_i\cdot\vec{\beta}>0}$, so the Fermat invariant is positive, and one accordingly gets from Eq.~\eqref{eq:omegap_rot} that ${\omega'<\omega<|\Omega_{ce}|}$. As a result, the wave index at a given location (i.e., density) is smaller, as illustrated in the right panel of Fig.~\ref{Fig:OX}. Because of the strong dispersion of the X mode in this region, the difference in wave index is noticeable even if $\beta$, and thus the Doppler shift, are small. Because, as mentioned above, the wave frequency moves away from the local resonance $\omega_{UH}$ as the density increases, dispersion decreases and the wave indexes with and without motion eventually match at lower radii. Nevertheless, the stronger index gradient at the edge in the presence of rotation leads to a stronger inward bending of the ray in this outer region. As shown in Fig.~\ref{Fig:OX}, this carries over to the reflected rays, which present different exit points and wavevectors with and without considering rotation. These findings suggest that accounting for motion effects could be important in X mode Doppler reflectometry modeling.

In short, these results show that motion affects the ray trajectory of waves in dispersive media through the Doppler shift it introduces, and underline that this effect may be non-negligible even for nonrelativistic velocities when dispersion is important.   

\subsubsection{Low-frequency wave drag from low group velocity}

To illustrate another manifestation of drag effect in this same rotating configuration, we now discuss briefly low-frequency waves ${\omega\leq\Omega_{ci}}$, for which only the X mode propagates in the form of the compressional Alfv{\'e}n wave. In this regime the wave index reduces to the simple form
\begin{equation} \label{eq:88a}
    n_X'\sim\sqrt{1+\left(\frac{c}{v_A}\right)^2}
\end{equation}
with ${v_A = c\Omega_{ci}/\omega_{pi}}$ the Alfv{\'e}n velocity. In contrast with the high-frequency case studied above, low-frequency propagation in the plasma rest-frame is thus nondispersive. However, it has been shown for a uniform motion that drag effects can manifest due to the large group index~\cite{langlois2025fresnel}.

\begin{figure}
    \centering
    \includegraphics{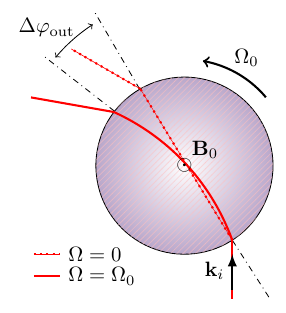}
    \caption{Ray trajectory of a low-frequency X wave in a rotating plasma cylinder verifying ${v_A/c\ll 1}$. In such a medium where ${n_g' \gg 1}$---i.e., a slow-light medium---the optical path is particularly sensitive to the effect of motion. This can be quantified by the angular deviation $\Delta\varphi_\mathrm{out}$ induced by motion, which is computed as the angular shift between the exit points with (solid line) and without (dotted line) motion.}
    \label{Fig:fig9}
\end{figure}

As a simple example, we consider as shown in Fig.~\ref{Fig:fig9} a solid body rotating plasma with uniform density. In this case and to lowest order in $\beta$, the ray trajectory Eq.~\eqref{eq:mpa} accounting for rotation can be rewritten in terms of this same trajectory without rotation, with 
\begin{equation} \label{eq:88b}
    \varphi(r)\sim\bar \varphi(r)+\frac{\Omega_0}{c}\int_r^R\frac{\xi n_X'^2}{\sqrt{\xi^2 n_X'^2-\mathfrak F_r^2}}{\rm d}\xi.
\end{equation}
In particular, this implies that the angular separation when exiting the rotating column verifies 
\begin{equation} \label{eq:klm1}
    \Delta\varphi_\mathrm{out} \defeq \varphi_\mathrm{out}-\bar \varphi_\mathrm{out} \geq 2\frac{\Omega_0}{c}\int_{\varrho}^R n_X'~{\rm d}\xi 
\end{equation}
where ${\varrho}$ is again the point of closest approach. If one further assumes an overdense plasma so that ${v_A/c\ll 1}$, then ${n_X'\gg1}$. The refracted wavevector is therefore nearly radial, implying that ${\varrho\rightarrow0}$. In this limit the angular deviation writes ${2\Omega_0 R/v_A}$, which one verifies is indeed the result obtained from the formula for transverse drag for a uniform linear motion derived by Player~\cite{player1975} if taking the transverse velocity at the edge $R\Omega_0$ as the linear velocity, ${v_A\ll c}$ as the group velocity, and $2R$ as the propagation length. 

Although very simplistic, this analysis shows that non-negligible ray deviations can occur for rays crossing a rotating plasma column if the tangential velocity is a fraction of the Alfv{\'e}n velocity, even if the angular velocity $\Omega_0$ of the plasma itself is relatively modest. Flows with Alfv{\'e}n Mach numbers approaching, or even greater than 1, are thus expected to produce large drag effects. Note finally that we simply considered for the sake of illustration a homogeneous plasma in solid body rotation, but generalization to more complex radial profiles of density, magnetic field and rotation are straightforward and captured in Eq.~\eqref{eq:mpa}.

\subsection{Nonuniform linear flow}

As a last illustration of drag in plasmas, we finally examine how rest-frame anisotropy and motion can compete with one another, considering the case where the phase and group velocities
in the comoving-frame are no longer collinear. 

\begin{figure}[htbp]
    \centering
    \includegraphics{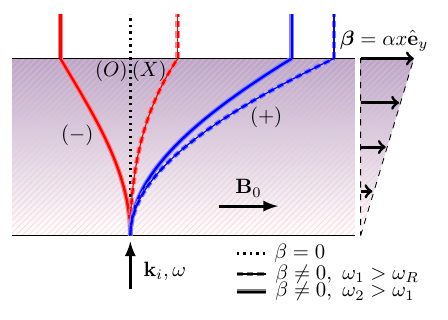}
    \caption{Wave at normal incidence on a magnetized plasma in nonuniform linear motion, with $\vec B_0$ in the incidence plane. The rays for the two refracted modes $(+)$ and $(-)$ are plotted for two different frequencies $\omega_1$ and $\omega_2$ as identified in Fig.~\ref{fig:linear_drag_plasma_freq}, illustrating that rays can be dragged in the same or in opposite directions.}
    \label{fig:linear_drag_plasma}
\end{figure}

For this let us go back to the case of a wave at normal incidence on a medium in linear nonuniform motion already examined in Section~\ref{sec:app}, but consider this time this medium to be a moving magnetized plasma with the magnetic field $\vec{B}_0$ aligned with the velocity, that is along the interface.  This is the configuration shown in Fig.~\ref{fig:linear_drag_plasma}, for which one verifies that ${\vec{B}_0=\vec{B}_0'}$. Because motion introduces a finite component along $\vec{B}_0$ to the rest-frame wavevector ${k_\parallel' = -\gamma\omega\beta/c}$, propagation in the rest-frame is oblique. Modes are then given by the biquartic Appleton-Hartree equation~\citep{bittencourt2013fundamentals}. Specifically, the normal component of the rest-frame wavevectors write
\begin{equation} \label{eq:solu}
    k_{\perp,\pm}' = \frac{c}{\omega'}\left[\frac{1}{2\Lambda}\left(-\Gamma \pm \sqrt{\Gamma^2-4\Lambda\Xi}\right) \right]^{1/2}.
\end{equation}
where
\begin{align} 
    \Lambda &= S, \\
    \Gamma &= \left(k_\parallel'c/\omega'\right)^2(P+S)-(LR+PS), \\
    \Xi &= P\left[\left(k_\parallel'c/\omega'\right)^2-L\right]\left[\left(k_\parallel'c/\omega'\right)^2-R\right],
\end{align}
and $P$, $L$, $R$, $S$ are the classical permittivities defined by Stix~\cite{stix1992waves}, which here are functions of the rest-frame frequency $\omega'$. Since $k_{\perp,\pm}'$ depends on the angle between the wavevector and the magnetic field, modeling drag requires the general equation Eq.~\eqref{eq:nonuniform_Psi}. Explicitly, one finds
\begin{equation}\label{eq:curv_eq_anisotrop2}
    y(x) = \int_x\gamma\Bigg\{c\beta \frac{\partial \mathcal K'}{\partial \omega'}(\omega', k_{\parallel}') - \frac{\partial \mathcal K'}{\partial k_{\parallel}'}(\omega', k_{\parallel}')\Bigg\} {\rm d}x
\end{equation}
where for normal incidence
\begin{align}
    \omega'(x)&=\gamma(x) \omega, \\
    k_{\parallel}'(x) &= -\gamma(x)\omega\beta(x)/c.\label{Eq:kprimepra}
\end{align}
Note that in the nonrelativistic limit ${\beta\ll1}$ the latter implies ${n_{\parallel}'\ll1}$. As a result propagation in the plasma is quasi-perpendicular other than when ${n'\rightarrow0}$, i.e., near cutoffs. 

\begin{figure}[htbp]
    \centering
    \includegraphics{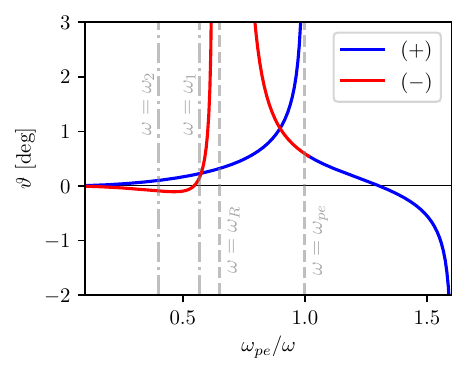}
    \caption{Drag angle $\vartheta$ as function of the lab-frame frequency $\omega$ computed for a fixed $\beta$ and normal incidence. Here $\beta=10^{-2}$, $B_0 = 1$~T and $N=10^{19}$~m$^{-3}$.}
    \label{fig:linear_drag_plasma_freq}
\end{figure}

Solving Eq.~\eqref{eq:curv_eq_anisotrop2}, Fig.~\ref{fig:linear_drag_plasma} shows that waves can be dragged in the direction of, or opposite to, the motion. To better understand how this behavior arises from the combination of anisotropy and motion, we consider drag as a function of lab-frame frequency, as plotted in Fig.~\ref{fig:linear_drag_plasma_freq}. Since, as mentioned above, $\vec{k}$ remains purely normal in the particular case of normal incidence, even for a nonuniform motion, we focus on the effect of a fixed velocity $\beta$, which can be thought of as the local $\beta$ in our nonuniform case. Specifically, we find in Fig.~\ref{fig:linear_drag_plasma_freq} that for a wave frequency much larger than the plasma natural frequencies $\Omega_{ce}$ and $\omega_{pe}$ the two modes are dragged in opposite directions (${\omega=\omega_2}$ in Fig.~\ref{fig:linear_drag_plasma_freq}), whereas they are both dragged along motion at lower frequency (${\omega=\omega_1}$ in Fig.~\ref{fig:linear_drag_plasma_freq}). Trying to shed light on this finding, we use that the generic drag formula Eq.~\eqref{eq:Psi} can be rewritten as 
\begin{equation}
    \tan{\vartheta} = \tan{\vartheta'}+\beta n_{g{\perp}}'.\label{Eq:varthetas}
\end{equation}
Here $\vartheta'$ is the angle between the normal to the interface and the group velocity in rest-frame, which here is different from the rest-frame refracted angle $\theta_t'$ since the phase and group velocities in the rest-frame are not aligned, and $n_{g\perp}'$ is the normal component of the rest-frame group-index. In the quasi-perpendicular regime of interest here, we know (see, e.g., Ref.~\cite{Booker1984}) that at high frequency the group velocity of the two modes are on either side of the wavevector, i.e., ${\vartheta_+'\vartheta_-'<0}$, with ${\vartheta_+'\rightarrow0}$ and ${\vartheta_-'\rightarrow0}$ when ${\omega'\rightarrow\infty}$. 
Since, from Eq.~\eqref{Eq:kprimepra}, the wavevector is itself inclined by an angle $\beta$ with respect to the normal, and ${n_{g{\perp}}'\rightarrow 1}$ for ${\omega\gg\omega_{pe}}$, Eq.~\eqref{Eq:varthetas} confirms that modes are dragged in opposite directions with ${\vartheta_-<0}$ and ${\vartheta_+>0}$, consistent with the behavior observed for ${\omega=\omega_2}$ in Figs.~\ref{fig:linear_drag_plasma} and \ref{fig:linear_drag_plasma_freq}. As the frequency decreases and approaches the generalized right cutoff $\omega_R$ (i.e., the frequency which reduces to the right cutoff of the X branch for purely perpendicular propagation) which for the parameters used here is ${\omega_R\sim\omega_{pe}/0.65}$, ${\vartheta_-'<0}$ decreases and approaches $-\pi/2$. Yet, at the same time ${n_{g_\parallel,-}'\rightarrow -\infty}$, which explains the drag reversal observed for the $(-)$ mode for ${\omega=\omega_1}$ in Figs.~\ref{fig:linear_drag_plasma} and \ref{fig:linear_drag_plasma_freq}. Indeed, rewriting Eq.~\eqref{Eq:varthetas} as ${\tan{\vartheta}  = \tan{\vartheta'}(1+\beta n_{g{\parallel}}')}$ shows that the drag direction will reverse for ${\beta n_{g{\parallel},-}'=-1}$. This result illustrates how rest-frame anisotropy and motion-induced drag can compete against each other.

Note finally for completeness that, similarly to the effect of dispersion exposed for isotropic media in Fig.~\ref{fig:linear_drag}, plasma motion can, in addition to the drag discussed just above, lead to turning points when the local Doppler-shifted frequency approaches a cutoff. For the particular case at hand the $(-)$ mode will, for instance, be reflected when the local Doppler-shifted, which is a decreasing function of $\beta$, approaches the plasma frequency $\omega_{pe}$. This effect of motion will be more marked for oblique incidence since motion then has a stronger effect on the rest-frame frequency. Furthermore, even if not leading to reflection in itself, motion could affect the turning point of a ray, for instance in a density gradient. Accounting for this effect could in principle, especially for sufficiently large $\beta$, be important for reflectometry diagnostics.

\section{Conclusions} \label{sec:conclu}

In this work we proposed a model for the effect of a nonuniform motion on the trajectory of a wave propagating in a moving anisotropic media. In this model the nonuniform flow appears in the form of spatial inhomogeneities in the properties (wave index) of the equivalent static medium, which then makes it possible to model rays as classically done in gradient index (GRIN) media. 

An evolution equation for the ray trajectory is first derived by considering how the local variation in velocity along the ray is the source of a Fresnel drag. This is analogous to the ray equation known in isotropic media, with the important difference though that this equation must here be complemented by an evolution equation for the wavevector since the ray direction and the wavevector are here not aligned.

It is then shown that symmetries in the velocity field translate into symmetries in the effective wave index, which then, analogously to GRIN media, can be used to obtain analytical solutions for the ray trajectory. Considering nonuniform linear motion and rotation, the predictions of this model are then compared with results previously derived for nondispersive isotropic media using an optical metric (also known as Gordon's metric), demonstrating the consistency of this new method with known results from the literature.

In contrast to the optical metric formalism, which has only been derived for isotropic media, the method proposed here applies to dispersive anisotropic media. To demonstrate this potential, the path of waves in a nonuniformly moving magnetized plasma is considered. The effect of motion on the classical O and X modes in a rotating magnetized plasma column is first exposed, showing that significant drag effects can occur  from dispersion, but also for waves with large the group index. Considering finally rays in a nonuniformly drifting plasma, for which propagation in the plasma rest-frame is oblique, it is found that, depending notably on the wave frequency, the rest-frame anisotropy of the plasma can add up or subtract to the effect of motion. Motion can therefore reverse the direction of refraction. It also affects the turning point condition of waves being reflected, for instance up a density gradient. For these reasons accounting for these effects could prove important for reflectometry measurements.

Besides applying these results to quantify motion effects on rays in practical applications, this work opens a number of interesting basic prospects. One lies in the possibility to use these results to derive an optical metric for dispersive anisotropic media. Such a tool would be particularly useful for numerical simulations. Another is the basic question of inertial corrections to rest-frame properties. In this work it was assumed, consistent with the literature, that the rest-frame properties are those of this same medium at rest. This, however, is only true for an inertial motion. For nonuniform motions the rest-frame constitutive relations are notably known to be modified~\cite{shiozawa1973phenomenological,langlois2023contribution}. Although these corrections have generally been deemed negligible~\cite{van1976electromagnetic}, the recent suggestion that they could in fact be dominant in unmagnetized plasmas~\cite{langlois2024signature} should motivate revisiting this problem. 

\begin{acknowledgments}
    This work was supported by the French Agence Nationale de la Recherche (ANR), under grant ANR-21-CE30-0002 (project WaRP). It has been carried out within the framework of the EUROfusion Consortium, via the Euratom Research and Training Programme (Grant Agreement No 101052200 – EUROfusion). Views and opinions expressed are however those of the authors only and do not necessarily reflect those of the European Union or the European Commission. 
\end{acknowledgments} 

\bibliography{references}

\end{document}